\documentclass[prl,two column]{revtex4}
\usepackage{amsmath}
\usepackage{graphicx}
\usepackage{xcolor}
\def\ra{\rangle}
\def\la{\langle}
\def\vr{\vec{r}}

\newcommand{\vs}{\vec{s}}

\newcommand{\STRUT}{\rule{0in}{3.0ex}}

\def\lsim{\mathrel{\raise3pt\hbox to 8pt{\raise -6pt\hbox{$\sim$}\hss{$<$}}}}
\def\ra{\rangle}
\def\la{\langle}

\begin{document}
\bibliographystyle{apsrev}

\title{Systematic Uncertainties in the Analysis of the Reactor Neutrino Anomaly}

\author{A.C. Hayes$^1$, J.L. Friar$^1$, G.T. Garvey$^1$, Gerard Jungman$^1$, G. Jonkmans$^2$}
\affiliation{$^1$Los Alamos National Laboratory, Los Alamos, NM, USA  87545\\
$^2$AECL, Chalk River Laboratories, Chalk River, Ontario, Canada, K0J 1J0}

\begin{abstract}

We examine uncertainties in the analysis of the reactor neutrino anomaly, wherein it is suggested that only about 94\% of the
emitted antineutrino flux was detected in short baseline experiments. We find that the form of
the corrections that lead to the anomaly are very uncertain for the 30\% of the flux that arises from
forbidden decays.
This uncertainty was estimated in four ways, is larger than the size of the anomaly, and is
unlikely to be reduced without accurate direct measurements of the antineutrino flux. 
Given the present lack of  detailed knowledge of the structure of the forbidden transitions, it is not possible to convert the measured
aggregate fission beta spectra to antineutrino spectra to the accuracy needed to infer an anomaly.  
Neutrino
physics conclusions based on the original anomaly need to be revisited, as do oscillation
analyses that assumed that the antineutrino flux is known to better than approximately $4\%$.        
\end{abstract}
\maketitle

The term ``reactor neutrino anomaly'' first appeared in a publication by G. Mention {\it et al.}
\cite{mention}, where it generally referred to the  $3\sigma$ deficit of neutrinos detected in short-baseline reactor neutrino experiments relative to the
number predicted.   The predicted number of detected neutrinos has evolved upward over time,
largely as a consequence of a predicted increase in the energy of the neutrino flux and an
increased  $\bar{\nu}_e + p \rightarrow n + e^+$ cross section associated with smaller values
for the neutron lifetime.  This cross section is used to infer the neutrino flux in a presumably
well-characterized detector.   The changes in the predicted neutrino flux are mostly associated
with improved knowledge of the beta decays of  the isotopes created in fission reactors.
Such an anomaly would potentially be  extremely significant, if a shortfall in the detected
neutrino flux could be ascribed to  $\bar{\nu}_e$ oscillation into a light sterile
neutrino with a mass of about 1~eV.  

There is an extensive recent literature dealing with the reactor anomaly, starting with a
seminal paper by Mueller {\it et al.} \cite{mueller}  that reexamined  the reactor antineutrino
flux.  The latter publication sought to improve the earlier flux estimates based on the ILL on-line
measurements \cite{schreck,schreck2,schreck3} of the integral beta spectrum of the fission products.  An
antineutrino spectrum can be inferred from a beta spectrum provided one knows the linear
combination of operators involved in the decay,  the end-point energy, and the nuclear charge. 
The fission beta spectra involve about 6000 beta transitions, of which about 1500 are
forbidden \cite{endf-7}.  Clearly some assumptions are required in order to infer the fission
antineutrino flux.
The improvements \cite{mention,mueller} on the earlier analyses of  ILL
integral measurements led to an increased energy of the antineutrino flux,
which was subsequently verified in an independent analysis \cite{huber}.

The present contribution examines the consequences of the
   forbidden transitions known to be present (at the 30\% level) in the beta decay
   of fission products.
We analyze the
antineutrino flux, using  a first-principles  derivation of the finite size (FS) and weak
magnetism (WM) corrections that were the main  focus of the analyses in Refs.~\cite{mueller,
huber}. In addition, we examine the shape factors needed to describe the forbidden
transitions.   We find that the forbidden transitions introduce a large uncertainty in the
predicted antineutrino flux  irrespective of whether the antineutrino spectra were deduced using
nuclear databases or by inverting measured aggregate fission beta spectra. As detailed below,
this finding results from the fact that the corrections are nuclear-operator dependent and that an
{\it undetermined} combination of matrix elements contributes to non-unique forbidden transitions. 
  
The beta-decay spectrum $S$ for a single transition  in nucleus $(Z,A)$ with end-point energy $E_0 =
E_e + E_\nu $ is 
\begin{equation}
S(E_e,Z,A) =  S_0 (E_e) F(E_e,Z,A) C(E_e) (1+\delta(E_e,Z,A)) \, ,
\label{allowed}
\end{equation}
where $S_0 = G_F^2 \, p_e E_e (E_0-E_e)^2 /2\pi^3$,
$E_e(p_e)$ is the electron total energy (momentum), $F(E_e,Z,A)$ is the Fermi function needed
to account for the Coulomb interaction
of the outgoing electron with the charge of the daughter nucleus, and $C(E_e)$ is a shape
factor  \cite{schopper} for forbidden transitions due to additional lepton momentum terms. For allowed transitions $C(E)=1$.  The
term $\delta(E_e,Z,A)$ represents fractional corrections to the spectrum that were the central
focus  of the original anomaly studies.  The primary corrections to beta decay are radiative,
finite size, and weak magnetism,
or $\delta(E_e,Z,A)=\delta_{\mathrm{rad}}+\delta_{\mathrm{FS}} +\delta_{\mathrm{WM}}$. 

Before discussing the details of the corrections $C(E_e)$ and $\delta(E_e)$, we briefly
summarize the treatments used in earlier work. The radiative corrections as derived by Sirlin
\cite{sirlin} were included in the description of the beta spectra (though not in the antineutrino
spectra) in the  original analyses of  Schreckenbach \emph{et al.} \cite{schreck,schreck2,schreck3}. In the
later ILL work \cite{schreck2,schreck3} an approximation for the FS and WM corrections was included by
first deducing the antineutrino spectrum from the measured beta spectra without these
corrections, and then applying a linear correction to the deduced antineutrino spectrum of the
form, $\delta_{\mathrm{FS}}+\delta_{\mathrm{WM}}=0.0065(E_\nu - {\rm 4\; MeV})$.  In that work
no corrections were
made for the shape factors $C(E_e)$. In the analyses of Refs.~\cite{mueller, huber, fallot} an
approximation (derived by Vogel \cite{vogel-2}) for the FS and WM corrections was applied on a
transition-by-transition basis.  In Refs.~\cite{mueller, fallot} the shape factor appropriate
for unique forbidden transitions was used for all forbidden transitions.  In Ref.~\cite{huber}
it was argued that these shape factors only play a small role in inferring antineutrino spectra
from measured beta spectra because for vanishing electron mass, $m_e$, they are symmetric under
$E_e \leftrightarrow E_\nu$.  In the present work,  we derived {\it ab initio} analytic
expressions for the FS and WM  corrections for allowed GT transitions, as well as WM and shape
factors for first-forbidden GT operators. We used the radiative corrections derived by Sirlin
\cite{sirlin}.

We now turn to the form of the corrections. The attractive Coulomb interaction {\it increases}
the electron density near the nuclear surface and increases the beta-decay rate, while the
finite  nuclear size {\it decreases} the electron density and decreases the rate (relative to
the point-nucleus Fermi function).  Using first-order perturbation theory in $Z\alpha$,  we find
that the finite-size correction to the Fermi function, $\delta_{\mathrm{FS}}$, for allowed GT transitions
is 
\begin{equation}
\delta_{\mathrm{FS}} = -\frac{3}{2}\frac{Z\alpha}{\hbar c}\left<r\right>_{(2)}\left(E_e - \frac{E_\nu}{27}+\frac{m_e^2c^4}{3E_e}\right)\\ 
\;\; .
\label{eqFS}
\end{equation}
The  quantity $ \la r \ra_{(2)}  = \int d^3r \, \rho_W (r) \int d^3 s \, \rho_{\rm ch} (s) \mid
\vr -\vs \mid$ is the first moment of the convoluted nuclear weak and  charge densities (called 
a Zemach moment \cite{zemach}).  We assume uniform distributions of radius $R$ for the weak and
charge densities, for which $\left<r\right>_{(2)}=\frac{36}{35}R$ \cite{friar1}.  The FS
corrections do depend on the beta-decay operator, but in this work we always use Eq.~(2) and $
R=1.2\;A^{1/3}$ fm.

The WM correction arises from the interference of the magnetic moment distribution of the vector
current, $\vec{J}_V=\vec{\nabla}\times\vec{\mu}$,  with the spin distribution $\vec{\Sigma}$ of
the axial current. We derived  the WM corrections for  allowed and first-forbidden operators.
There are four possible operators in the case of first-forbidden GT transitions, and all have
well-defined WM corrections, as listed  in Table 1.  
\begin{table*}
\caption{The shape factors and leading-order weak magnetism corrections to allowed and first-forbidden
Gamow-Teller beta decays are shown in the top panel. The shape factors for allowed and first-forbidden
Fermi beta decays are shown in the bottom panel. All agree with Ref.~\cite{milliner} for $Z=0$.
The entries for $\vec{J}_V$ and $\rho_A$ are discussed in  \cite{siegert}.
The weak magnetism correction for $\vec{J}_V$ involves the unknown overlap of very different $1^-$ matrix
elements and is therefore not listed. The nucleon isovector magnetic moment is $\mu_v =4.7$, $M_N$
is the nucleon mass, $g_A$ is the axial vector coupling constant, and $\beta = p_e/E_e$. No meson currents 
were used in the magnetic moment operator, and a truncated orbital current led to the factor of ``1/2'' in $\delta_{\mathrm{WM}}$.
}
\vspace{4pt}
\begin{tabular}{ccccc}
Classification & $\;\Delta J^\pi\;$ & \; Operator \; & Shape Factor $C(E_e)$ & \; Fractional Weak Magnetism Correction $\delta_{\mathrm{WM}}(E_e)$\\\hline\hline
Allowed GT&$1^+$&$\Sigma\equiv\sigma\tau$&1&$\frac{2}{3}\left[\frac{\mu_v-1/2}{M_Ng_A}\right](E_e\beta^2-E_\nu)$\\
Non-unique 1$^{st}$ Forbidden GT&$0^-$&$\left[\Sigma,r\right]^{0-}$&$p_e^2+E_\nu^2+2\beta^2E_\nu E_e$&0\\
\!\!Non-unique 1$^{st}$ Forbidden $\rho_A$ &$0^-$&$\left[\Sigma,r\right]^{0-}$&$\lambda\,  E_0^2$&0\\
Non-unique 1$^{st}$ Forbidden GT&$1^-$&$\left[\Sigma,r\right]^{1-}$&$p_e^2+E_\nu^2-\frac{4}{3}\beta^2E_\nu E_e$&$\;\;\;\;\left[\frac{\mu_v-1/2}{M_Ng_A}\right]\left[\frac{(p_e^2+E_\nu^2)(\beta^2 E_e-E\nu)+2 \beta^2 E_e E_\nu(E_\nu-E_e)/3}{(p_e^2+E_\nu^2-4\beta^2E_\nu E_e/3)}\right]$\\
Unique 1$^{st}$ Forbidden GT&$2^-$&$\left[\Sigma,r\right]^{2-}$&$p_e^2+E_\nu^2$&\,\,$\frac{3}{5}\left[\frac{\mu_v-1/2}{M_Ng_A}\right]\left[\frac{(p_e^2+E_\nu^2)(\beta^2 E_e-E\nu)+2 \beta^2 E_e E_\nu(E_\nu-E_e)/3}{(p_e^2+E_\nu^2)}\right]$\STRUT \\[1.5ex] \hline 
Allowed F &$0^+$&$\tau$&1&0\\
\!\!\!Non-unique 1$^{st}$ Forbidden F&$1^-$&$r\tau$& $p_e^2+E_\nu^2+\frac{2}{3}\beta^2E_\nu E_e$& 0\\[0.7ex] 
Non-unique 1$^{st}$ Forbidden $\vec{J}_V$ &$1^-$&$r\tau$& $E_0^2$& -\\[0.7ex]
\hline
\end{tabular}
\end{table*}
Our FS and WM corrections for allowed GT transitions are identical to those derived by Holstein
\cite{holstein}, and differ from the forms used in \cite{mueller, huber, fallot, vogel-2}.   The
first-forbidden shape factors, $C(E_e)$, which depend on the operator in question, were derived
and  are displayed  in Table 1. To determine the implications of the  corrections we take
advantage of the fact that the ENDF/B-VII.1 decay library \cite{endf-7} contains branching
ratios  and end-point energies for  over 90\% of the transitions making up the full spectra. 
Thus, the library allows a representative estimate of  the effect of the corrections, provided
we know what assumptions to make about the forbidden transitions.

We first fit the Schreckenbach\cite{schreck2} electron spectrum for $^{235}$U before examining the effect of the corrections in an analysis based on the summation of the known beta branches of the fission products. We include the corrections both with and without a treatment of the forbidden transitions. 
We fit the beta spectrum with 40 fictitious equally spaced end-point energies using non-negative least-squares fitting \cite{nnls}.
The charge Z associated with each end-point is assigned according to the method given in \cite{schreck2}.
There is no unique physical prescription for beta-decay operator assignments to the fictitious end-points.
For this reason we examine four prescriptions: (1) all transitions are assumed to be allowed; (2) all end-point energies can be associated with either an allowed or forbidden transition;
  (3) 30\% of the branches are selected to be forbidden at  equal energy intervals; (4)  30\% of the branches are selected to be forbidden with a bias towards higher energies. 
In addition, we examine fits in which the operator determining the forbidden decays was taken to be $[\Sigma,r]^{0-}, [\Sigma,r]^{1-}$, $[\Sigma,r]^{2-}$ or a combination of these.
As a natural consequence of the non-negative least-squares procedure, the fit results in a significant fraction of the end-points having zero amplitude. The subset of end-points with non-zero amplitudes 
varies depending on the operator assignment.  

We find excellent fits to the electron spectrum in  all cases. 
However, different treatments of the forbidden transitions can lead to antineutrino spectra that differ
both in shape and magnitude at about the 4\% level.
Two examples are shown in Fig.1, where we present the fits obtained when the WM and FS corrections 
are included. In one case all transitions are assumed to be allowed, while in the second case the best fit results from about 25\% forbidden decays.
For the assumption of all allowed transitions, we see a systematic increase  of about 2.2\% in the number of antineutrinos relative to Schreckenbach, while including forbidden transitions leads to  no increase relative to Schreckenbach.
Other prescriptions for the forbidden transitions lead to changes relative to the Schreckenbach antineutrino spectrum ranging from
0-4\%, including changes in the shape of the antineutrino spectrum.
{\sl These examples help to clarify the difficulty in inferring with high certainty the 
antineutrino spectrum from a measured electron spectrum when information on the forbidden
transitions is not available.}
\begin{figure}
\includegraphics[width=3.5 in]{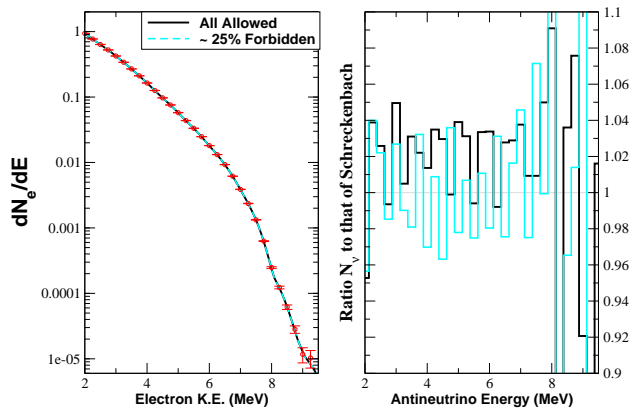}
\vspace*{-0.3in}
\caption{ The fit to the  electron spectrum for $^{235}$U (left) for two different assumptions on how to treat forbidden transitions, and the ratio of the corresponding  antineutrino spectra to that of Schreckenbach (right). 
The electron data are those of \cite{schreck2}.
The electron spectra are fit assuming (a) all allowed GT branches, or (b) up to 30\% forbidden GT transitions. In both cases the WM and FS corrections are included. 
When folded over the neutrino detection cross section, the case for all allowed (25\% forbidden) transitions results in
a 2.2\% (0.06\%) increase in the number of detectable antineutrinos.
} 
\label{fig:fit}
\end{figure}

In calculating the aggregate fission spectra from the database library an analogous uncertainty 
arises because detailed  structure information does not exist for the majority of the  roughly 1500
forbidden transitions.
In addition, several of the transitions are forbidden at second or  higher order, for which
analytic corrections are not available.
Thus, there is no clear prescription for applying the corrections to this component of the
spectra, and it is more beneficial to examine the effect of different approximations in order to
estimate the uncertainty involved. 
In all approximations we treat unique forbidden transitions as unique first-forbidden GT
transitions, and treat non-unique forbidden transitions in one of the following ways: (1) as
allowed GT;  (2) as unique first-forbidden GT with the operator
$\left[\Sigma,r\right]^{2-}$;  (3) with the operator
$\left[\Sigma,r\right]^{0-}$; (4) with the operator $\left[\Sigma,r\right]^{1-}$.
None of the these  treatments is correct, 
but they provide estimates for changes in the spectra induced by forbidden transitions.

The aggregate fission beta spectrum under equilibrium reactor burning conditions for a given
actinide is determined by the beta spectra $S(E_e,Z_i,A_i)$ of the individual unstable fission
fragments weighted by their cumulative fission yields, $Y_{F_i}$ \cite{tal}:
\begin{equation}
N_\beta(E_e) = \sum_{F_i} Y_{F_i} S(E_e,Z_i,A_i)  .
\label{agg}
\end{equation}
The beta spectrum $S$ for each fragment $(Z_i,A_i)$ summed over all decay branches  must be
normalized to unity: $\int S(E,Z,A)\,dE=1$.  
Thus, Eq.~(\ref{agg}) is a statement that under equilibrium burning conditions the beta-decay
rates are determined by the fission rate \cite{nonequilibrium}.
If the antineutrino spectrum is inferred from a measured aggregate beta spectrum,  Eq.~
(\ref{agg}) must be replaced by a sum over a set of end-point energies $\{E_{0_i}\}$, weighted
by a fitted set of coefficients $\{a_i\}$: $N_\beta=\sum_i a_i S(E_e,E_{0_i})$. 

\begin{figure}[h]
\includegraphics[width=3.5 in]{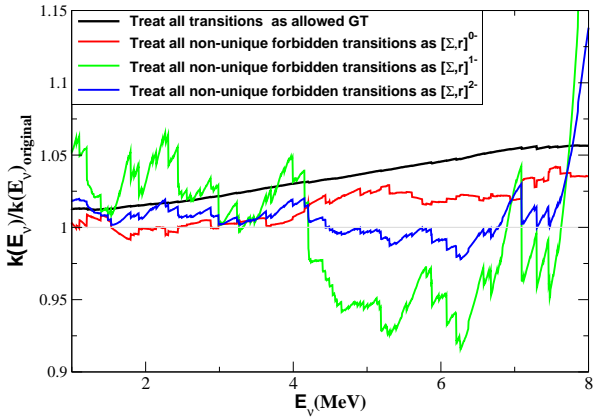}
\caption{ The ratio of the function $k(E_e,E_\nu)$ for $^{235}$U using ENDFB/VII.1 relative to using $E_\nu=K_{\beta}$ \cite{borovoi}.}
\label{fig:ke-function}
\end{figure}

There is no unique method for determining the uncertainty in the antineutrino spectrum
introduced by the forbidden transitions.
Another possibility (in addition to the fitting exercise above) is to consider changes in the
bi-variant function $k(E_e,E_\nu)$,  where $k(E_e,E_\nu)=N_\nu(E_\nu)/N_\beta(E_e)$.  
If $k(E_e,E_\nu)$ only changes by some small percentage for some path in the $(E_e,E_\nu)$ plane
as we change our treatment of forbidden transitions,  then there exists a prescription for
inferring the antineutrino spectrum to that accuracy.  
We calculated the function $k(E_e,E_\nu)$ for each of our four assumptions (above). 
We found {\it no path} in the $(E_e,E_\nu)$ plane that left $k(E_e,E_\nu)$ unchanged by as
little as 5\% as our assumptions for the forbidden transitions changed. 
Fig.~(\ref{fig:ke-function}) depicts the result for the path $E_\nu=K_{\beta} \equiv E_e - m_e c^2$ \cite{borovoi}.
Similar or larger differences were found for all other paths. 
The very non-smooth and non-linear shape of $k$ arises from the shape factors $C(E_e)$. 

A third prescription for estimating the uncertainties is to examine the rate of change in the
antineutrino spectrum relative to the rate of change in the beta spectrum, using the fact  that
the beta spectrum is fit to  amplitudes ${a_i}$ on a fixed grid of  end-point energies
${E_{0_i}}$. We calculated  $T= \sum_i [\partial N_\nu(E_\nu)/\partial a_i ]/[ \partial
N_\beta(E_e)/ \partial a_i] $, and  examined the changes in $T$ as the assumptions for the
forbidden transitions  were varied, and again found no path in $(E_e,E_\nu)$ over which the 
changes were $<$ 5\%.
\begin{figure}[h]
\includegraphics[width=3.8 in]{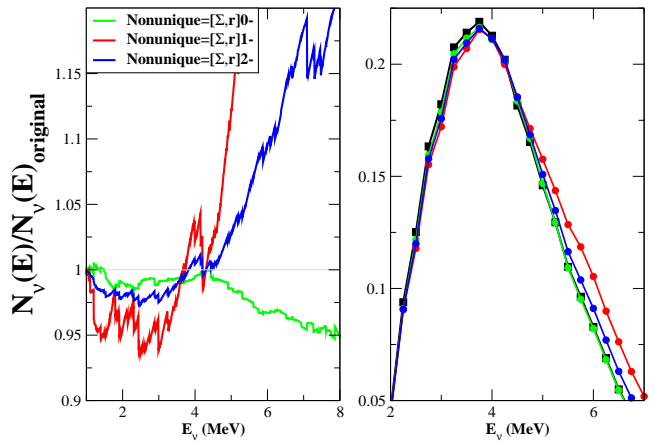}
\caption{Different treatments of the forbidden GT transitions contributing to the antineutrino spectrum
summed over all actinides in the fission burn in mid-cycle \cite{kopeikin} of a typical reactor.
The left panel shows the ratio of these antineutrino spectra relative to that using the assumptions
of Ref.~\cite{schreck2}.
The right panel shows the spectra weighted by the detection cross section, where the additional curve
in black uses the assumptions of Ref.~\cite{schreck2}.
The spectra are strongly distorted by the forbidden operators, being lower below the peak  and in some
cases more than 20\% larger above the peak  than Ref.~\cite{schreck2}.
The corresponding change in the number of detectable antineutrinos relative to \cite{schreck2} is -0.75\%, 5.8\% and 1.85\% for
the $0^-, 1^-$, and $2^-$ forbidden operators, respectively.}
\label{fig:ke-neut}
\end{figure}

Our final method considers the ratio of the actual antineutrino spectra themselves in Fig.~(\ref{fig:ke-neut}),
although this method does not take into account the corresponding changes in the beta spectra.
We carried out identical analyses of the role of the corrections and the associated
uncertainties for the other actinides $^{239,241}$Pu and $^{238}$U, and found very similar
results. Fig.~(\ref{fig:ke-neut}) shows the change in the total antineutrino spectrum at a representative time
\cite{kopeikin}  in  mid-cycle in the reactor burn history.  The antineutrino spectra differ
significantly, depending on our treatment of the forbidden transitions. The
cross-section-weighted spectra are quite distorted, being lower than the Schreckenbach-Vogel
\cite{schreck2,vogel-1,explain} spectra up to the peak, and higher or lower above the peak
depending on the operator.  The actual spectrum is unlikely to be as distorted as in Fig.~(\ref{fig:ke-neut})
because no single operator dominates the forbidden transitions.

The original ILL analysis \cite{schreck2} assumed that the $Z$ of the daughter fragments
satisfies $Z_i=49.5-0.7E_{0_i}-0.09E_{0_i}^2 $ for $Z_i\ge 34$. We find that taking the $Z$ of
the fission fragments directly from ENDF/B-VII.1 instead  would increase  the antineutrino
spectrum by less than 1\% for $E_\nu \leq$ 7 MeV, and less than 1.5\% for $E_\nu >$ 7 MeV. 

In summary we find that the component of the  aggregate fission
spectra containing approximately 30\% forbidden transitions introduces a large uncertainty (about 4\%) in the predicted shape of the
antineutrino flux emitted from reactors.
 We have examined the uncertainties in four different
ways.  If all forbidden transitions are treated as allowed GT transitions, the antineutrino
spectra are systematically increased above 2 MeV, as was the conclusion in the earlier papers on
the anomaly. However, when the forbidden transitions are treated in various approximations, the
shape and magnitude of the spectra are changed significantly. 
Earlier analyses only looked at one prescription for these transitions.
The uncertainty introduced by our
lack of knowledge on how to treat these transitions is as large as the size of the anomaly. 
It should also be noted that there are additional uncertainties due to our oversimplified treatment of 
the allowed weak magnetism operator (viz., no meson currents and a truncated orbital current), as well as a 
simplified finite-size treatment for forbidden transitions. These considerations also apply to fission antineutrino fluxes in  medium and long baseline reactor experiments, implying a 4\% uncertainty on the fission antineutrino flux in those experiments. Reducing the uncertainty within
a purely theoretical framework would be difficult.  An improvement will require either {\it direct}
measurements of the antineutrino flux or a substantial improvement in our knowledge of the dominant forbidden beta transitions.


\begin{thebibliography}{99}
\bibitem{siegert} Nuclear operators $\vec{J}_V$ and $\rho_A$ are proportional to a nucleon velocity ($p/M_N$), have negative parity, and are therefore first-forbidden. Replacing them by the analogous operators proportional to $E_0\, r$ requires only CVC for $\vec{J}_V$,  but is an approximation for $\rho_A$ made exact by introducing an unknown dimensionless nuclear-structure parameter $\lambda$ that differs from case to case.
\bibitem{nonequilibrium}Off-equilibrium corrections are needed for fission fragments that decay with half-lives long compared to the timescales involved in the experiments in \cite{schreck,schreck2,schreck3}.
\bibitem{explain} ``Schreckenbach-Vogel''  means that we used the spectra for $^{235}$U, $^{239}$Pu, and $^{241}$Pu from \cite{schreck2} and $^{238}$U from \cite{vogel-1}.
\end{thebibliography}
\end{document}